\documentclass[letterpaper,12pt]{article}

\usepackage{tabularx} % extra features for tabular environment
\usepackage{authblk}  % for authors and affiliations
\usepackage{amsmath}  % improve math presentation
\usepackage{graphicx} % takes care of graphic including machinery
\usepackage[margin=1in,letterpaper]{geometry}  % decreases margins
\usepackage[superscript,biblabel,nomove]{cite} % takes care of citations
\usepackage[final]{hyperref} % adds hyper links inside the generated pdf file
\usepackage{setspace} % for double spacing
\usepackage{paralist} % for itemize without spacing
\hypersetup{
	colorlinks=true,       % false: boxed links; true: colored links
	linkcolor=blue,        % color of internal links
	citecolor=blue,        % color of links to bibliography
	filecolor=magenta,     % color of file links
	urlcolor=blue         
}
\title{Risk-based decision making: estimands for sequential prediction under interventions}

\author[1]{Kim Luijken}
\author[2,3]{Pawe{\l} Morzywo{\l}ek}
\author[1]{Wouter van Amsterdam}
\author[4,5,6]{Giovanni Cinà}
\author[7]{Jeroen Hoogland}
\author[8]{Ruth Keogh}
\author[9]{Jesse Krijthe}
\author[10]{Sara Magliacane}
\author[11]{Thijs van Ommen}
\author[12]{Niels Peek}
\author[13]{Hein Putter}
\author[1]{Maarten van Smeden}
\author[12]{Matthew Sperrin}
\author[14]{Junfeng Wang}
\author[14]{Daniala Weir}
\author[15,16]{Vanessa Didelez}
\author[* 13]{Nan van Geloven}

\affil[*]{Corresponding author, \url{n.van_geloven@lumc.nl}}
\affil[1]{Department of Epidemiology, Julius Center for Health Sciences and Primary Care, University Medical Center Utrecht, Utrecht, the Netherlands}
\affil[2]{Department of Applied Mathematics, Computer Science and Statistics, Ghent University, Ghent, Belgium}
\affil[3]{Department of Statistics, University of Washington, Seattle, United States}
\affil[4]{Department of Medical Informatics, Amsterdam University Medical Centers, Amsterdam, the Netherlands}
\affil[5]{Institute for Logic, Language and Computation, University of Amsterdam, Amsterdam, the Netherlands}
\affil[6]{Pacmed, Amsterdam, the Netherlands}
\affil[7]{Department of Epidemiology and Data Science, Amsterdam University Medical Centers, Amsterdam, the Netherlands}
\affil[8]{Department of Medical Statistics, London School of Hygiene \& Tropical Medicine, Keppel Street, London, United Kingdom}
\affil[9]{Pattern Recognition and Bio-Informatics Group, EEMCS, Delft University of Technology, Delft, the Netherlands}
\affil[10]{Amsterdam Machine Learning Lab, University of Amsterdam, Amsterdam, the Netherlands}
\affil[11]{Department of Information and Computing Sciences, Utrecht University, Utrecht, the Netherlands}
\affil[12]{Division of Informatics, Imaging and Data Science, Faculty of Biology, Medicine and Health, University of Manchester, Manchester Academic Health Science Centre, Manchester, United Kingdom}
\affil[13]{Department of Biomedical Data Sciences, Leiden University Medical Center, Leiden, the Netherlands}
\affil[14]{Division of Pharmacoepidemiology and Clinical Pharmacology, Department of Pharmaceutical Sciences, Utrecht University, Utrecht, the Netherlands}
\affil[15]{Department of Biometry and Data Management, Leibniz Institute for Prevention Research and Epidemiology - BIPS, Bremen, Germany}
\affil[16]{Faculty of Mathematics/Computer Science, University of Bremen, Bremen, Germany}

\date{\today}

\begin{document}

\maketitle

\noindent The authors declare no conflict of interest and state that no additional data was used.

\begin{doublespacing}

\section*{Risk-based decision making: estimands for sequential prediction under interventions}
\begin{abstract}
Prediction models are used amongst others to inform medical decisions on interventions. Typically, individuals with high risks of adverse outcomes are advised to undergo an intervention while those at low risk are advised to refrain from it. Standard prediction models do not always provide risks that are relevant to inform such decisions: e.g., an individual may be estimated to be at low risk because similar individuals in the past received an intervention which lowered their risk. Therefore, prediction models supporting decisions should target risks belonging to defined intervention strategies. Previous works on prediction under interventions assumed that the prediction model was used only at one time point to make an intervention decision. In clinical practice, intervention decisions are rarely made only once: they might be repeated, deferred and re-evaluated. This requires estimated risks under interventions that can be reconsidered at several potential decision moments. In the current work, we highlight key considerations for formulating estimands in sequential prediction under interventions that can inform such intervention decisions. We illustrate these considerations by giving examples of estimands for a case study about choosing between vaginal delivery and cesarean section for women giving birth. Our formalization of prediction tasks in a sequential, causal, and estimand context provides guidance for future studies to ensure that the right question is answered and appropriate causal estimation approaches are chosen to develop sequential prediction models that can inform intervention decisions.
\end{abstract}

\noindent \textbf{Keywords:} Counterfactual prediction - Estimand - Prediction model - Prediction under interventions

\section{Introduction}
To enhance health and health care, there is a need for smart decision support tools that can improve medical decision making by providing care professionals and patients with person-specific information. Clinical prediction models estimate risks of (future) outcomes conditional on patient characteristics and thus have the potential to provide such information\cite{moons2015transparent, moons2009prognosis, steyerberg2009clinical, riley2019prognosis}. Yet, standard prediction models are generally not suitable for supporting intervention decisions. Prediction models that are intended to inform decisions about interventions need to answer questions like `what is the risk of outcome $Y$ under intervention option $a$ conditional on an individual's characteristics $\mathbf{X}$?'. Answering such `what if' questions requires embedding in causal reasoning\cite{sperrin2018using, sperrin2019explicit,van2020prediction,dickerman2020counterfactual,dickerman2022predicting,coston2020counterfactual,lin2021scoping,paper2.1}. We refer to this task as `prediction under (hypothetical) interventions'\footnote{Prediction under interventions has also been referred to as `counterfactual prediction’\cite{hernan2019second, dickerman2020counterfactual, coston2020counterfactual}. We use the term prediction under interventions instead because at the moment of making the prediction the outcomes are still in the future and hence risks need not be counterfactual\cite{dawid2010identifying}.}.\par

A number of studies have emphasized the need for prediction under interventions and clarified how to estimate risks under interventions\cite{sperrin2018using, sperrin2019explicit,van2020prediction,dickerman2020counterfactual,dickerman2022predicting,coston2020counterfactual,lin2021scoping}. The focus of these studies was on intervention decisions that were made at a single time point, but in clinical practice intervention decisions are rarely made only once and might often be deferred or re-evaluated. To support such sequential decisions, prediction models are needed that can estimate risks under interventions at any potential decision moment. By sequentially providing estimated outcome risks under intervention options conditional on individual characteristics, the model functions as an \textit{assistive} decision support tool\cite{reilly2006translating}. Methods for sequential predictions are well established in a standard prediction context, but not yet for interventional predictions\cite{rizopoulos2012joint, van2011dynamic}. The current work clarifies how to formulate and formalize sequential predictions under interventions.\par

The topic of the current work, \textit{sequential prediction under interventions}, is different from the related topic of optimal dynamic intervention regimes. A dynamic intervention regime is a function that takes baseline covariates, covariate history, and intervention history as inputs and returns an intervention decision to be made next\cite{chakraborty2013statistical, moodie2007demystifying, murphy2003optimal, robins2004optimal, dawid2010identifying}. Methods exist for deriving optimal dynamic intervention regimes which are optimal with respect to a predefined utility function. Optimal dynamic intervention rules provide advice on what intervention to assign at a moment in time. Methods for deriving optimal dynamic decision rules assume that an evaluation function exists to assess the value of candidate decision rules. In practice, the challenge is often to come up with such a utility function: it will determine everything. Our work can provide utility functions – in particular, those that assess the risk of adverse health outcomes under a given candidate decision rule. Simply providing information on outcome risks under interventions can sometimes be more useful to end users than a suggested intervention decision. This is because patients and care professionals can have different utility functions and their value judgements and view on what is optimal can change over time. An optimal intervention rule requires assignment of an intervention as recommended by the decision rule during all stages in order to arrive at the optimal outcome. Care professionals may be hesitant to consistently follow an algorithm's suggestions for multiple intervention decisions ahead. Instead, having estimated risks of outcomes under intervention options on the table can facilitate a conversation between care professional and patient so that informed decisions can be made\cite{phd_pawel,van2020prediction}.\par

The starting point of development of a prediction model under interventions is formulating the risk questions that reflect the desired predictions, that is, formulating the estimands. The current work highlights key considerations for formulating estimands for sequential prediction under interventions. We illustrate these considerations by giving examples of estimands for single-stage and sequential prediction under interventions in a case study about deciding on vaginal delivery or cesarean section for women giving birth. The contribution of this paper is to define prediction tasks in a sequential, causal, and estimand context. A clearly defined estimand is needed to ensure answering the right question and choosing appropriate causal estimation approaches.

\section{How to formulate estimands for prediction under interventions}
We start with providing general considerations for formulating estimands for prediction under interventions and then highlight aspects that require additional attention in the sequential setting.\par

An estimand is a precise definition of the target quantity of an analysis. In prediction studies, an estimand captures the conditional outcome risk (or value) of interest that is intended to be be used as information about a particular individual. The conditional outcome risk can be interpreted as \textit{individualized} rather than individual, because they are formalized at a group level given a covariate pattern\cite{hoogland2021tutorial, knaus2021machine}. The term `estimand' is rarely used in prediction modelling studies, but prediction guidelines implicitly provide recommendations on aspects that should be defined to specify the target of a prediction study\cite{moons2014critical,moons2015transparent, whittle2017prognosis, luijken2019impact}. By combining these recommendations with roadmaps available from clinical trial literature\cite{ich2020addendum} and causal inference literature\cite{petersen2014causal, goetghebeur2020formulating}, we arrive at the following elements that are required for estimands for prediction under interventions:
\begin{itemize}
    \item Population: A characterisation of the target population to whom the prediction model is to be applied, including the care setting;
    \item Moment(s) of intended use: The moment(s) at which the prediction model is to be used to inform the intervention decision;
    \item Intervention options: The interventions considered at the moment(s) of intended use and, if relevant, for how long they are administered;
    \item Outcome and prediction horizon: The predicted outcome including the time since the moment of prediction at which we consider it. The specified outcome and prediction horizon should represent information that is needed in the discussion among care professionals and patients to make the intervention decision;
    \item Predictor(s): The measurement(s) used for prediction that characterize the individual, i.e., that individualize the predictions. The measurement(s) must be available at the moment(s) of intended use and measurement procedures must correspond to those in the setting of intended use.
\end{itemize}

\subsection{Considerations for prediction under interventions at a single time point}

As an example of a model that provides predictions under interventions at a single time point, an estimand could express the risk of a binary outcome $Y\in\{0,1\}$ at a prediction horizon $h$, conditional on predictors $\mathbf{X}$ measured at moment of intended use under specified intervention options $A\in\{0,1\}$ in a given population. Introducing notation, we denote regular time points by $k = 0, 1, 2, \ldots, K$. Observed intervention status $A_k$, predictors $\mathbf{X}_k$, and outcome status $Y_k$ are measured at each time point. We assume the moment of intended use is time point $k = 0$. We let $\underline{a}_{0} = ( a_0, \dots, a_{K} )$ be a specified joint intervention, setting the intervention to a particular level $a_k$ for the times between $0$ and $K$. Here, $\underline{a}_{0} = \mathbf{0}$ indicates not initiating the intervention at the moment of prediction and continuing to do so until time point $K$ and $\underline{a}_{0} = \mathbf{1}$ indicates initiating the intervention at the moment of prediction and continuing to do so until time point $K$. An intervention option might also refer to a well-defined standard of care. We define the potential outcome $Y_h^{\underline{a}_0}$ as the outcome at prediction horizon $h$ if an individual is assigned intervention sequence $\underline{a}_0$. By fixing the intervention to a certain strategy, an estimand for prediction under intervention at a single time point can be formally defined as:
\begin{equation*}
    \Pr{\left[Y^{\underline{a}_{0}}_h=1|\mathbf{X_0}\right]},
    \label{estimand_example}
\end{equation*}
e.g., for $\underline{a}_{0}=\mathbf{0}$ this would be the conditional risk of outcome $Y$ at prediction horizon $h$ for an individual with predictor values $\mathbf{X_0}$ if they would not initiate the intervention up to the prediction horizon.\par

To precisely define the intervention options for the estimand, it is helpful to make a distinction between different types of interventions. Point interventions are administered once or for a (very) short duration, like single-dose pharmacological treatments or surgery. Sustained interventions are administered during longer disease episodes and can take the form of a \textit{static} or \textit{dynamic} regime\cite{sterne2016robins,chakraborty2013statistical}. A static regime specifies a fixed sequence of the intervention status for a pre-defined duration. For instance, ``an individual starts the intervention of interest and continues to use it during a year". Static regimes can be seen as joint interventions as they do not take information into account that occurs during the course of time. Dynamic regimes were mentioned in the introduction and propose an intervention decision based on the (time-varying) health state of an individual. For instance, ``an individual starts the intervention of interest when marker X drops below a certain value". Point interventions, static intervention regimes, and dynamic intervention regimes can all be used as intervention options in an estimand for prediction under interventions.

\subsection{Considerations for sequential prediction under interventions}\label{general_sequential}
When an intervention decision can be deferred or re-evaluated, a prediction model capable of sequentially estimating risks under hypothetical interventions is needed. \par 

We consider a setting in which predictions of a binary outcome are made at regular time points $k = 0, 1, 2, \ldots$. The prediction horizon can either stay fixed for each $k$, i.e., $h_k=h$, or can be defined relative to the moment of prediction, e.g., $h_k = k+ w$. Let $\underline{a}_{k}= ( a_k, \dots, a_{K} )$ be the specified intervention sequence of interest. We let ${\mathbf{ \overline{X}}}_{k}=(\mathbf{X}_0, \mathbf{X}_1, \mathbf{X}_2, \ldots, \mathbf{X}_{k})$ denote the observed predictor history up to $k$, ${\overline{A}}_{k}=({A}_0, {A}_1, {A}_2, \ldots, {A}_{k})$ denote the observed intervention history up to $k$, and $Y_{k}$ denote whether the outcome has occurred before time $k$. An estimand for sequential prediction under an intervention option can then be defined as:
\begin{equation*}
    \Pr{\left[Y^{\underline{a}_{k}}_{h_k}=1 | \overline{\mathbf{{X} }}_{k} , Y_{k}=0, {\overline{A}}_{k}\right]},
    \label{estimand_example_seq}
\end{equation*}
e.g., for $\underline{a}_{k} = \mathbf{0}$ and $\overline{A}_{k} = \mathbf{0}$, this estimand reflects the conditional outcome risk at time $k+w$ for an individual with a predictor history $\mathbf{\overline{X}}_{k}$ who did not experience the outcome or initiate the intervention before time $k$, if they would not initiate the intervention up to the prediction horizon by deciding on `no intervention' each time the need for intervening is re-evaluated.

An estimand for sequential prediction under interventions incorporates updated predictor information. The time until the next prediction moment informs how long the intervention options could be fixed and at what time point the prediction horizon could be set. For instance, if prediction moments follow each other at short intervals, it might be more reasonable to define a short-term prediction horizon and to fix the intervention option to a certain level until that prediction horizon. This is likely less reasonable if predictions under interventions are made infrequently. An overview of key considerations regarding estimands for sequential prediction under interventions  is given in Table~\ref{tab:elements}.\par

\begin{table}[h!]
    \begin{tabularx}{\textwidth}{lX}
    \hline
    \textbf{Estimand element} & \textbf{Questions that help in formulating the estimand element} \\
    \hline
    Population & \begin{compactitem}
        \item To which individuals will the prediction model be applied?
        \item In which health care setting will the prediction model be applied?
    \end{compactitem} \\
    Moment(s) of intended use & \begin{compactitem}
        \item At which moment(s) is the prediction model (re)consulted to inform the intervention decision?
    \end{compactitem} \\
    Intervention options & \begin{compactitem}
        \item Which intervention options are relevant at the moment(s) of making the intervention decision?
        \item For how long should the intervention strategy be fixed?
        \item Should the duration to fix the intervention option be aligned with the time till next moment of prediction?
    \end{compactitem} \\
    Outcome and prediction horizon & \begin{compactitem}
        \item Which outcome(s) are most informative for the intervention decision?
        \item What prediction horizon provides important information for the intervention decision: a short-term or long-term horizon?
        \item Should the outcome be defined differently because of the specified intervention option(s)?
        \item Should the prediction horizon be aligned with the time till next moment of prediction?
    \end{compactitem} \\
    Predictor(s) & \begin{compactitem}
        \item Which predictors are predictive of the outcome of interest?
        \item Based on which characteristics should the outcome risks be individualized?
        \item Which measurements are available at the moment(s) of intended use?
    \end{compactitem} \\
    \hline
    \end{tabularx}
    \caption{Considerations to define estimands for sequential prediction under interventions.}
    \label{tab:elements}
\end{table}

\newpage

\section{Case study: mode of delivery in women with high-risk pregnancies}
We use a clinical case study to illustrate the required considerations for formulating estimands for sequential prediction under interventions. We define several example estimands for a prediction model that informs the intervention decision for mode of delivery in women giving birth. We start out with a setting in which the prediction model is only used once to inform the decision. For this single-stage prediction, we formulate four example estimands that vary in their intervention options. The example is then extended to sequential prediction under interventions with three example estimands that vary in intervention options and prediction horizon.

\subsection{Clinical context}
The clinical case study is inspired by a prediction model developed by Schuit and colleagues that was proposed to support medical decisions during labor of high-risk pregnant women\cite{schuit2012clinical}. Different from the original model, in the current study we define the outcome as a composite of adverse events in the mother and child. To simplify the discussion, we focus on the interventions vaginal delivery and cesarean section only, and implicitly include other interventions, like ways to induce natural vaginal delivery or instrumental vaginal delivery, under vaginal delivery.\par

As a brief background on the clinical setting based on Schuit and colleagues\cite{schuit2012clinical}; some pregnancies are classified as `high-risk' because they are complicated by pre-existing maternal disease or complications during pregnancy. High-risk pregnant women are typically monitored by gynaecologists in secondary care and are admitted to hospital to give birth. Birth can take place via different modes of delivery including natural vaginal delivery and cesarean section. Cesarean section can be a way to prevent adverse neonatal outcomes due to fetal distress, but it comes with risks for the mother like increased blood loss, incontinence, and infection\cite{NICE2021cesarean}. The decision to give birth via cesarean section can be made before start of labor so that a planned cesarean section can be performed at an elected time\cite{NICE2019highrisk}. When high-risk pregnant women not scheduled for planned cesarean section go into labor, the decision about mode of delivery is re-evaluated at start of labor and repeatedly during labor. The decision to perform cesarean section is based on a weighting of benefits and harms of the procedure and preferences of care professionals and the woman giving birth\cite{NICE2023intrapartum, NICE2019highrisk}.\par

\subsection{Elements shared across the example estimands}
Common elements of the estimands formulated in the case study are:
\begin{itemize}
    \item Population: Pregnant women at a gestational age of 36 weeks or over with preexisting maternal disease or complications during pregnancy admitted to hospital to give birth;
    \item Moment(s) of intended use: At the start of labor for the single-stage prediction and hourly from start of labor for the sequential prediction;
    \item Intervention options: Intervention options considered are vaginal delivery and cesarean section, where the duration of fixing the intervention varies across estimands;
    \item Outcome and prediction horizon: The outcome is defined as a composite of any adverse neonatal or maternal outcomes. Adverse neonatal outcomes include stillbirth, early neonatal death, or requiring admission to a neonatal intensive care unit. Adverse maternal outcomes include maternal death and postpartum admission to intensive care. The prediction horizon varies across example estimands;
    \item Predictors: Fetal heart rate, dilatation, maternal systolic blood pressure, maternal diastolic blood pressure, maternal age, parity, and history of preterm birth. For the sequential prediction, we additionally use hourly-updated information of the time-varying predictors (fetal heart rate, dilatation, maternal systolic blood pressure, and maternal diastolic blood pressure).
\end{itemize}

\subsection{Single-stage prediction under interventions}
\begin{figure}[!b]
    \centering
    \includegraphics[width = \textwidth]{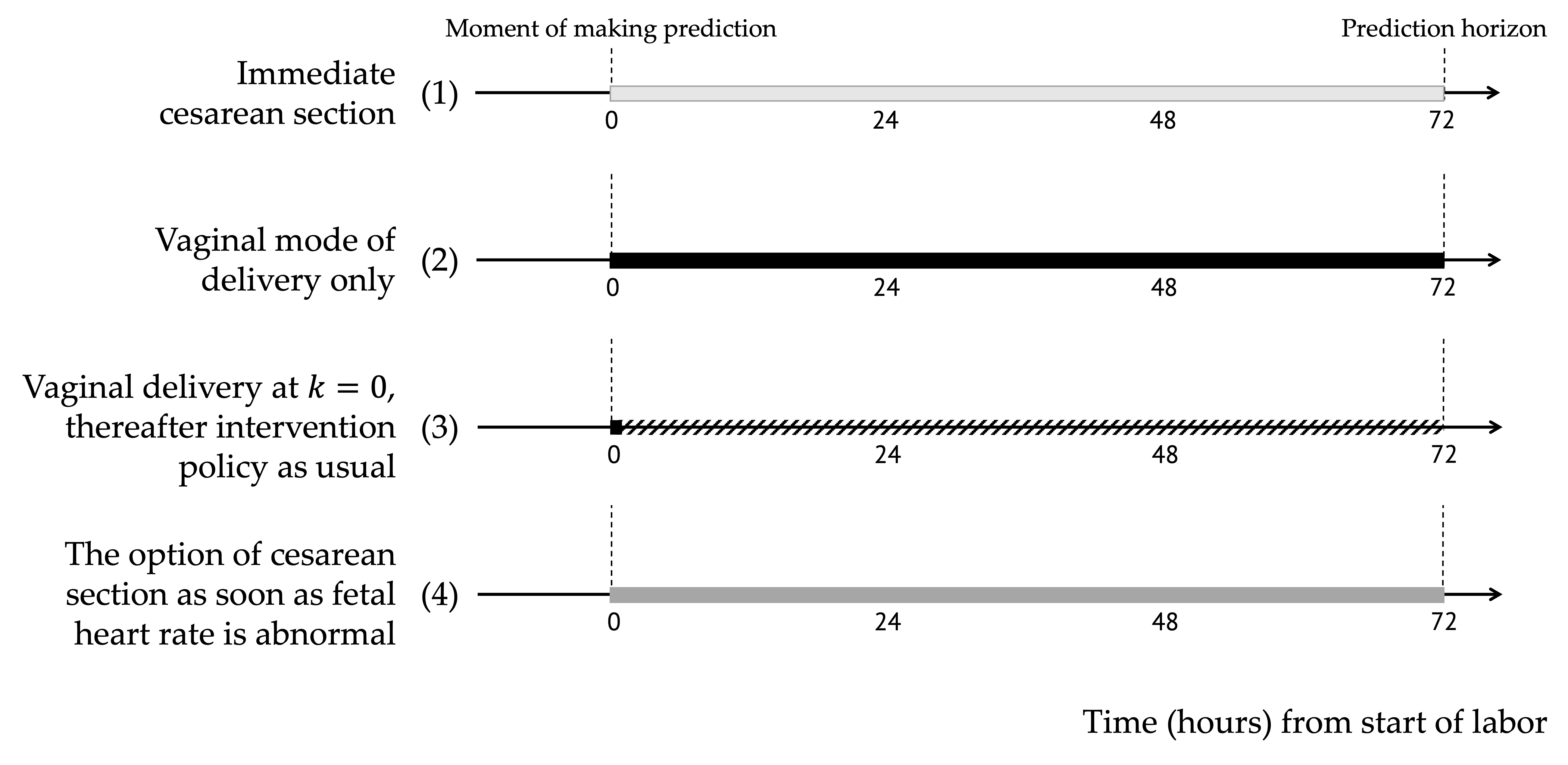}
    \caption{Illustrations of estimands for single-stage prediction under intervention options in the case study. The shaded rectangles indicate the intervention options under which predictions are made: (1) Light grey for cesarean section at time point 0, which is irreversible; (2) Black for vaginal mode of delivery only; (3) Black for setting mode of delivery to vaginal delivery at time point 0, and diagonal stripes for the option of cesarean section as in usual care; (4) Dark grey for the dynamic intervention rule `performing a cesarean section as soon as fetal heart rate is abnormal'.}
    \label{fig:est1to4}
\end{figure}

In the single-stage prediction setting, we assume the prediction model is used only at the start of labor to estimate risks of adverse outcomes 72 hours later under different intervention options. We vary the defined intervention options across four example estimands (Figure \ref{fig:est1to4}). Note that a single-stage prediction assumes a single time point at which predictions are made to inform the intervention decision, but these predictions can apply to a sequence of (hypothetical) future interventions.

\newpage
\subsubsection*{Notation}
We introduce some further notation specific to the case study. Let $k = 0, 1, 2, \ldots, 72$ be time points of one hour apart, where $k = 0$ denotes the start of labor and $h=72$ denotes the prediction horizon. Let $\underline{a}_{0}=(a_0, a_1, a_2, \ldots, a_{71})$ denote an intervention option for the 72 time intervals, where $a_k \in \{0,1\}$, $a_k=1$ denotes cesarean section at time $k$ and $a_k=0$ denotes deciding on vaginal mode of delivery at time $k$. Because cesarean section is irreversible, $a_k$ stays $1$ after it was set to $1$ for the first time by the nature of the intervention. We use $\underline{A}_{k}=(A_k, \ldots, A_{72})$ to denote a vector of observed intervention status between time $k$ and time $72$. Let $\mathbf{X_0}$ denote the set of seven predictors available at the start of labor: fetal heart rate, dilatation, maternal systolic blood pressure, maternal diastolic blood pressure, maternal age, parity, and history of preterm birth. Let $Y_{72}\in\{0,1\}$ denote the outcome status at the prediction horizon, where $Y_{72}=1$ represents any of the defined adverse neonatal or maternal outcomes having occurred and $Y_{72}=0$ represents completion of birth in absence of the adverse outcomes. Because all women can be expected to have given birth after 72 hours, the chosen prediction horizon implies that the outcome reflects all adverse events.\par 

\subsubsection*{Estimand 1: Immediate cesarean section}
To formulate the outcome risk when cesarean section is performed immediately at the start of labor, we define the intervention option as setting mode of delivery to cesarean section at $k=0$, i.e., ${a}_0=1$. Then,
\begin{equation}
    \Pr{\left[Y^{{a_{0}}=1}_{72}=1|\mathbf{X_0}\right]}
    \label{estimand_ces}
\end{equation}
expresses the conditional risk of adverse neonatal or maternal outcomes 72 hours after start of labor under immediate cesarean section at start of labor given the predictors at start of labor. Note that predictors after time point 0 cannot be considered, because this would imply that model users would need to input information about the future at moment of intended use.\par

\subsubsection*{Estimand 2: Vaginal mode of delivery only}
Another intervention option would be a sustained static intervention regime from the start of labor until the prediction horizon, e.g., $\underline{a}_{0}=\mathbf{0}$. Then,
\begin{equation}
    \Pr{\left[Y^{\underline{a}_{0}=\mathbf{0}}_{72}=1|\mathbf{X_0}\right]}
    \label{estimand_vagallcost}
\end{equation}
expresses the conditional risk of adverse neonatal or maternal outcomes 72 hours after start of labor under vaginal mode of delivery only, given the predictors at start of labor.

\subsubsection*{Estimand 3: Deciding on vaginal mode of delivery at moment of prediction and thereafter intervention policy ``as usual"}
Estimand (\ref{estimand_vagallcost}) informs the intervention decision by predicting outcomes under perfect adherence to the intervention. The woman in labor and health care professionals may prefer a vaginal delivery but the intervention strategy `vaginal mode of delivery only' likely does not reflect actual practice. In practice, women and care professionals re-evaluate the need for intervening and may decide to change the mode of delivery. Information on the conditional risk of adverse events under practice as usual can be obtained, for example, by setting an intervention option only for the moment at which the prediction is made, without committing to a particular intervention choice or decision rule at future time points. Predicting under usual practice refers to the typical practice and decision rules used in the development/training data.\par

To estimate such risks, we define the intervention option as setting mode of delivery to vaginal delivery between $k = 0$ and $k=1$, i.e., $a_{0}=0$, and leaving the option of choosing cesarean section later if indicated according to usual intervention assignment during all subsequent times $k$, i.e., $\underline{A}_{1}$\cite{phd_pawel}$^{\text{, Ch.~5}}$. This implies that $a_0$ is fixed to a value, whereas $\underline{A}_{1}$ is a random variable taking on the "natural" values, i.e. what would happen under the policy used in the development/training data. Then,

\begin{equation}
    \Pr{\left[Y^{a_{0}=0, {\underline{ A}_{1}} }_{72}=1|\mathbf{X_0}\right]}
    \label{estimand_vagusual}
\end{equation}
expresses the conditional risk of adverse neonatal or maternal outcomes 72 hours after the start of labor under deciding on vaginal mode of delivery for the first hour after start of labor and the option of cesarean section at subsequent time points according to usual intervention policy, given the predictors at start of labor. The way that the intervention option is specified in Estimand (\ref{estimand_vagusual}) relates closely to the way this is done in the intervention contrast of an intention to treat effect.\par

The appeal of Estimand (\ref{estimand_vagusual}) is that the intervention option is similar to the way a decision might be made in clinical practice: an intervention option is now chosen, but there is no commitment to a particular intervention option or decision rule at future points in time. The downside of this is that the conditional outcome risks are dependent on the intervention assignment that was observed in the development/training data. This implies that estimated risks will only generalize to settings with similar intervention assignment policies\cite{van2020prediction}. Predictions under a predefined static intervention or (optimal) dynamic intervention do not rely on this assumption.\par

\subsubsection*{Estimand 4: Performing a cesarean section as soon as fetal heart rate is abnormal}
At the start of labour, a possible decision is to perform a cesarean section as soon as but only if there are signs of fetal hear rate abnormality. More specifically, we define a dynamic intervention rule that cesarean section is performed the first time fetal heart rate is less than 110 beats per minute for 3 minutes (persistent fetal bradycardia) or higher than 160 beats per minute (fetal tachycardia). This is an example of a basic dynamic intervention rule, in which an intervention is assigned based on the history of a single covariate up to that stage. More generally, dynamic intervention rules can also be based on multiple covariates such as taking the mother`s health status into account. It could be an optimal dynamic rule as described in the introduction as well.\par

Let $X_k$ denote fetal heart rate at time point $k$ and $g_k$ an indicator function that flags abnormal fetal heart rate. 

\begin{equation*}
    g_k( {x}_k) = 
    \begin{cases}
      0 & \text{if $110 \leq x_k \leq 160$}\\
      1 & \text{if $x_k < 110 \text{ for 3 minutes in a row  or } x_k > 160$}
    \end{cases} 
\end{equation*}

The dynamic intervention sequence $d$ equals 0 until the first time point $k$ where $g_k(X_k) = 1$, and 1 from that point on wards. As the $X_k$'s are random, the sequence $d$ is random as well, we only fix the rule on how to respond to variations in $X_k$. The estimand of interest can be specified as
\begin{equation}
    \Pr{\left[Y^{\underline{a}_0=d}_{72}=1 |\mathbf{X}_0\right]}.
    \label{estimand_vagdynam}
\end{equation}

This estimand expresses the conditional risk of adverse neonatal or maternal outcomes under the intervention option to perform cesarean section as soon as the fetal heart rate is abnormal, given the predictors at start of labor. \par

It deserves clarification as to why we consider a dynamic intervention strategy for the single-stage prediction setting, which may be a confusing classification. The dynamic intervention rule states that based on fetal heart rate during the course of labor, cesarean section can be performed. This means that the need for intervening is re-evaluated over time. However, the choice to \textit{commit to this rule} is made only once, at the start of labor. In other words, the action (of starting cesarean section) is re-evaluated, the decision is not. To support the decision, we only need a single-stage prediction of the conditional outcome risk under the dynamic intervention rule (Figure~\ref{fig:est1to4}-4). Next, we will consider estimands where the prediction under intervention options is re-evaluated over time.

\subsection{Sequential prediction under interventions}
Predictions under interventions can be considered several times during the course of labor to inform the decision on mode of delivery. We continue the example by assuming that the prediction model is used at the start of labor and thereafter revisited every hour to update the risks. When defining an estimand that is suitable for sequential prediction under interventions, it is important to consider the time-varying information up to the moment of making the prediction as well as the time axes for the intervention options and outcome after the moment of making the prediction. We discuss these considerations using three examples of relevant estimands. 

\subsubsection*{Further notation}
We now assume that predictions to inform the decision about mode of delivery are made every hour $k = 0, 1, 2, \ldots$. We keep the prediction horizon at 72 hours after start of labor. In other settings, the prediction horizon may be defined relative to the moment of making the prediction, as explained in Section~\ref{general_sequential}. The predictors $\mathbf{X}_k$ are measured at each time point\footnote{We realize that maternal age, parity, and history of preterm birth are time-fixed predictors, but do not explicate this in the notation for simplicity.}. Let ${ \overline{\mathbf{X }}}_{k} =(\mathbf{X}_0, \mathbf{X}_1, \mathbf{X}_2, \ldots, \mathbf{X}_{k})$ denote the predictor history up to time point $k$. We denote whether the woman has given birth at time $k$ by introducing an `at risk' indicator $Z_k\in \{0,1\}$, where $Z_k = 1$ denotes that the woman is still in labor and no adverse outcome has occurred yet and $Z_k = 0$ denotes that the woman has given birth or an adverse outcome has occurred. 

\subsubsection*{Estimand 5: Vaginal mode of delivery only, using updated predictor information}
We first extend Estimand (\ref{estimand_vagallcost}) to a sequential prediction setting using updated information at the moment of prediction. The starting point is to consider for which individuals the intervention decision still is to be made, i.e., which individuals form the `risk set' at time $k$. In this example, the intervention decision on mode of delivery is to be made in women who are still in labor and who have not experienced an adverse event. We use $Z_k=1$ to select the risk set of women who are still in labor at time $k$.\par

\begin{figure}[!b]
    \centering
    \includegraphics[width = \textwidth]{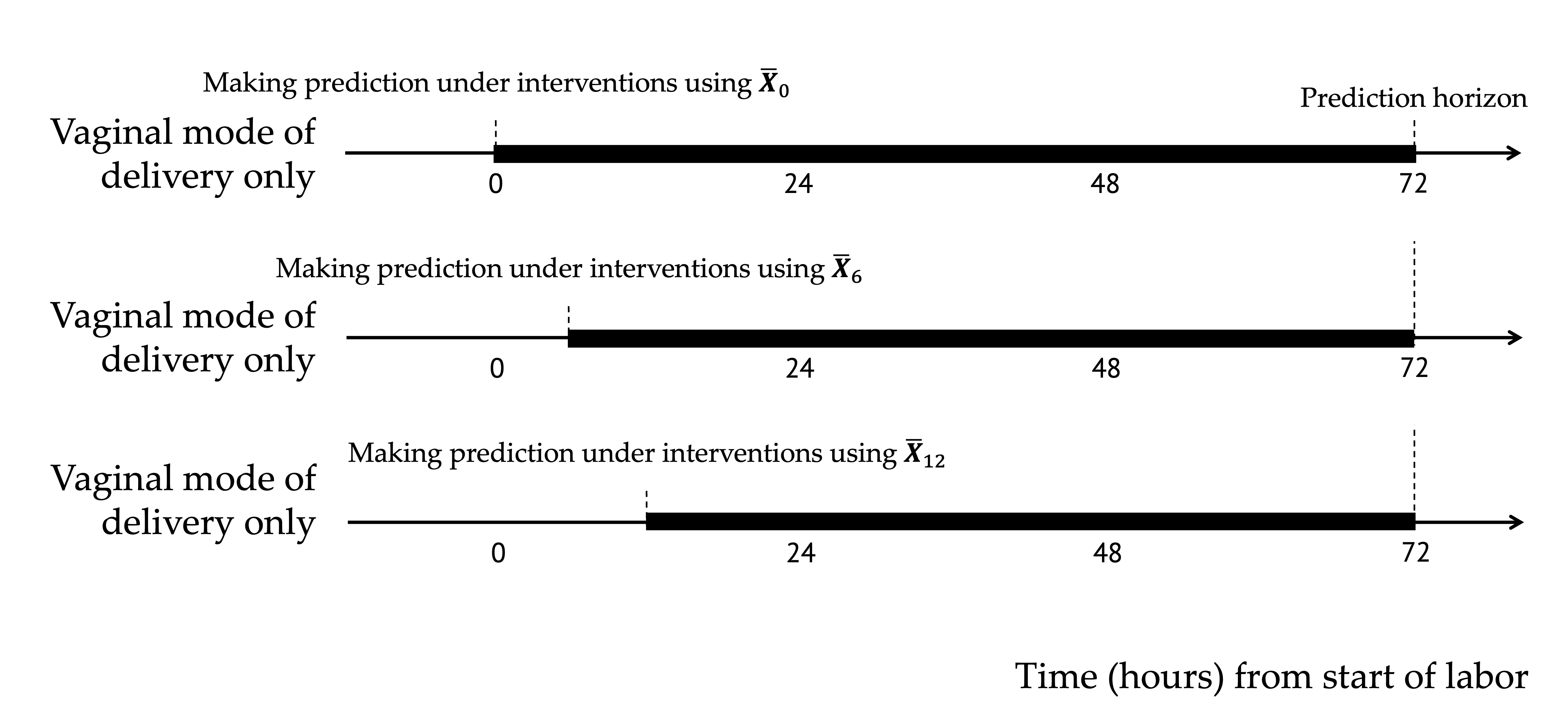}
    \caption{Depiction of sequential prediction under interventions for Estimand 5. The black rectangles indicate the intervention option under which predictions are made: vaginal delivery, which is sustained throughout labour. The moment of making the prediction and prediction horizon shift over time to provide sequential updated predictions.}
    \label{fig:est5}
\end{figure}

Predictor information up to the moment of making the prediction is $\overline{\mathbf{X }}_{k}$. We define the intervention option as setting mode of delivery to vaginal delivery from time $k$ onwards for the remaining time of labor, i.e., $\underline{a}_{k}=\mathbf{0}$. Then,

\begin{equation}
    \Pr{\left[Y^{\underline{a}_{k}=\mathbf{0}}_{72}=1| \overline{\mathbf{X }}_{k}, Z_k=1\right]}
    \label{estimand_vagallcost_seq}
\end{equation}
expresses the conditional risk of adverse neonatal or maternal outcomes 72 hours after start of labor under vaginal mode of delivery only given the predictor history up to time $k$ and that the woman is still in labor and no adverse outcome has occurred at time $k$ (Figure \ref{fig:est5}). We retrieve Estimand~\ref{estimand_vagallcost} as the special case where the moment of intended use if restricted to $k = 0$.

\newpage
\subsubsection*{Estimand 6: Vaginal mode of delivery at moment of prediction and thereafter intervention policy ``as usual"}
A consideration in sequential prediction is to define the duration of fixing the intervention strategy under which predictions are made. As explained in the discussion of Estimand (\ref{estimand_vagusual}), fixing the intervention strategy for the entire prediction window has benefits in terms of generalizability, but it might not be an intervention option that corresponds closely to actual clinical practice. In the current example, it might be preferable to fix the intervention for one hour because the prediction model is revisited every hour.\par

Define the intervention option as setting mode of delivery to vaginal delivery for one hour at time $k$, i.e., $a_{k}=0$, and follow usual intervention policy during all subsequent times, i.e., $\underline{A}_{k+1}$. Then,

\begin{equation}
    \Pr{\left[Y^{a_{k}=0, {\underline{A}_{k+1}} }_{72}=1| \overline{\mathbf{X }}_{k}, Z_k=1\right]}
        \label{estimand_vagusual_seq}
\end{equation}
expresses the conditional risk of adverse neonatal or maternal outcomes 72 hours after start of labor under vaginal delivery for the next hour and having the option of cesarean section at subsequent time points according to usual intervention policy. This risk is conditional on the predictor history up to time $k$ and that the woman is still in labor and no adverse outcome has occurred at time $k$. The duration over which the intervention option is fixed is in this case informed by the time till next decision moment. We retrieve Estimand~\ref{estimand_vagusual} as the special case where the moment of intended use if restricted to $k = 0$.\par 

\subsubsection*{Estimand 7: Vaginal mode of delivery at moment of prediction and a short prediction horizon}
Finally, estimands for sequential prediction require that the time axis of the outcome, i.e., the prediction horizon, is defined. This element of the estimand determines what information the prediction model provides to inform the intervention decision. In the example, the prediction model is used to estimate risks of relatively acute outcomes (within 72 hours) for the mother and child. In other cases a long prediction horizon may be apt, for instance when predicting (side) effects that might develop months or years later. When the predictions are revisited regularly, it might be preferable to estimate a conditional outcome risk until the next moment the prediction model is consulted. We could for example define the prediction horizon at $k+1$. Then,

\begin{equation}
    \Pr{\left[Y^{a_{k}=0}_{k+1}=1| \overline{\mathbf{X }}_{k}, Z_k=1\right]}
    \label{estimand_vagallcost_seqshort}
\end{equation}
expresses the conditional risk of adverse neonatal or maternal outcomes within one hour after consulting the prediction model under vaginal delivery for the next hour given the predictor history up to time $k$ and that the woman is still in labor and no adverse outcome has occurred at time $k$. The duration of fixing the intervention and the prediction horizon are in this case informed by the time till next decision moment. Because of the short prediction horizon, it might also be informative to define an intermediate outcome instead, such as fetal hypoxia.

\section{Remarks on identification and estimation}
After estimands for prediction under interventions are defined in a study, the subsequent steps are assessing identifiability given the observed data and choosing an appropriate estimator. Identifiability and estimation are beyond the scope of the current work, but we provide a few pointers here.\par

Assessing whether the observed information is sufficient to identify the desired estimand for prediction under interventions involves evaluating whether causal identification assumptions are met. The identifiability assumptions typically include consistency, (conditional and sequential) exchangeability, and (sequential) positivity\cite{hernan_causal_2020,boyer2023assessing,keogh2023prediction}. We refer the reader to other material for discussions on how to assess identifiability, such as \cite{hernan_causal_2020, boyer2023assessing,keogh2023prediction}, but make two remarks relevant to prediction under interventions in particular. The predictor(s) that are defined as part of an estimand are typically chosen based on (bedside) availability and prognostic value. They are not selected with causal assumptions in mind. Evaluation of (sequential) exchangeability might reveal that additional (time-varying) covariates need to be taken into account during estimation for confounding adjustment. Regarding (sequential) positivity, it needs to be possible to observe individuals under all formulated intervention options for all covariate patterns given the predictors in the model. This is more stringent compared to (sequential) positivity for average treatment effects and often implies that more data is needed or parametric assumptions need to be made. Alternatively, one could take a pragmatic approach and focus on intervention options that are identifiable given the available data\cite{keogh2023prediction}.\par

If an estimand is identifiable given observed data, we can then select an estimator from many available existing methods to estimate the formulated estimands, mainly stemming from causal inference literature on dynamic/conditional treatment effect estimation\cite{chakraborty2013statistical, morzywolek2023general, gran2010sequential}. Such methods can be used to develop a prediction model under the hypothetical intervention scenario(s) as if all individuals had followed the defined intervention option\cite{van2020prediction, sperrin2018using, lin2021scoping, dickerman2022predicting}. In the sequential prediction context, additional care should be given to the way the history of the predictors up to the moment of prediction is summarized\cite{rizopoulos2012joint, van2011dynamic, putter2022landmarking, rizopoulos2023using}. Further methodological work is needed to develop dedicated methods for estimating and validation of prediction under interventions\cite{keogh2023prediction}, for example combining the selection of predictors from a set of candidate predictors while remaining sufficient adjustment for confounding.

\section{Discussion}
The current work provides recommendations about how to formulate estimands for sequential prediction under interventions. We combined guidance for formulating estimands in prediction, clinical trial and causal inference literature and discussed key considerations (Table~\ref{tab:elements}). The considerations were illustrated by formalizing seven estimands in a clinical case study about the decision on mode of delivery in women giving birth. These insights on estimands extend previous work on predictions under interventions for single-stage decisions and link principles of dynamic prediction modelling to prediction under intervention options.\par

In empirical studies focusing on sequential prediction under interventions, it is likely that multiple estimands are relevant for the prediction problem at hand. How to align the moments of intended use, intervention duration, and prediction horizon should depend on context-specific knowledge and preferences of different model users. Ultimately, the selected (set of) estimand(s) should capture the identifiable risks under intervention options that provide relevant information for decision making. The choice of moments of intended use of a prediction model under interventions can also be determined by estimating the optimal timing of risk assessments\cite{gasperoni2023optimal}.\par

We discussed estimands for an example of an assistive prediction model that can provide information on risks for mother and child under the defined intervention options. These risks are one piece of information that could inform the decision on mode of delivery and leaves weighting of other sources of information to arrive at a decision to the model users. As such, \textit{assistive} prediction models fit the process of shared decision making more naturally compared to optimal dynamic intervention rules (which are sometimes referred to as \textit{directive} decision support tools\cite{reilly2006translating}).\par

The transition from outcome risk estimation to a particular intervention or triage decision is typically described to take place only after model development and validation, in a so-called impact analysis study\cite{reilly2006translating, moons2009impact}. In an impact analysis, the impact of the use of a prediction model on patient outcomes or (cost) efficiency of care is evaluated in a setting in which some healthcare professionals are assigned to use the prediction model when making decisions about an intervention -- preferably by randomization. The current work links predictions to decisions already in the model development phase, which ensures that predictions are aligned to their intended clinical impact. We recommend that development of such prediction models always starts by defining relevant estimands. The need for other good practices in prediction modelling, like external validation and impact studies, remains\cite{de2022guidelines, keogh2023prediction, boyer2023assessing}.\par

Causal research closely related to prediction under interventions focuses on estimation of individualized intervention \textit{effects} conditional on covariates with the goal to investigate heterogeneity in intervention effects in randomized controlled trials\cite{hoogland2021tutorial, kent2020predictive, kent2020predictive-ee,kent2018personalized,cai2022relate}. Still, most work in this area seems to assume that an intervention decision is made at a single point in time, at the moment of randomization, which often does not align with decision making in clinical practice. Furthermore, there is debate whether a conditional difference in means should be used to assign interventions based on who benefits most\cite{vanderweele2019selecting, penning2020re}. Our work focuses on individualized outcome risks under certain intervention options rather than estimating intervention contrasts so that the prediction model can be used as an assistive tool in making an intervention decision.\par

Our clinical case study did not address all intricacies of decisions to be made when defining estimands for sequential prediction under interventions. For example, others have proposed to set the intervention for the first time point and afterwards assume an optimal intervention regime\cite{vansteelandt2014structural}. The use case contained some specific features that do not occur in general such as irreversibility of the intervention `cesarean section'. We specified a composite outcome, but knowing the risk of each outcome separately might be more helpful in practice to weigh the risks for the mother and neonate. The prediction horizon was set at a time point of 72 hours, which was relevant in the clinical context, but may complicate the estimation when a single model is used for prediction windows of different duration. Because our focus was on defining estimands, we kept the discussion of estimation strategies to a minimum. Yet, estimating conditional outcome risks under interventions might be particularly challenging in a sequential setting.\par	

Defining a suitable estimand for sequential prediction under interventions is far from trivial, but is a pivotal starting point for development of any prediction model intended to inform intervention decisions at multiple time points. Different estimands can produce risk estimates that are relatively similar, but in some situations estimates can differ substantially\cite{van2020prediction,prosepe2022disconnect}. Formulating estimands prevents intervention decisions from being misguided by information from prediction models. The current work illustrates how estimands for sequential prediction under interventions can be formulated. 

\newpage

\bibliographystyle{unsrt}
\bibliography{references.bib}

\newpage
\subsection*{Conflict of interest}
The authors declare no conflict of interest.

\subsection*{Data availability statement}
The authors state that no additional data was used for this study.

\subsection*{Funding}
No funding was received for this study.

\end{doublespacing}
\end{document}